\documentclass[prl,aps,twocolumn,showpacs,superscriptaddress]{revtex4}
\usepackage{graphicx, amsmath, amsthm, subfigure}
\makeatletter
\begin{document}
\title{Local tunneling spectroscopy as signatures of the
Fulde-Ferrell-Larkin-Ovchinnikov state in $s$- and $d$-wave superconductors}
\author{Qian Wang}
\affiliation{Texas Center for Superconductivity, University of Houston, Houston,
Texas 77204}
\author{H.-Y. Chen}
\affiliation{Texas Center for Superconductivity, University of Houston, Houston,
Texas 77204}
\author{C.-R. Hu}
\affiliation{Department of Physics, Texas A\& M University, College Station,
Texas 77843}
\author{C. S. Ting}
\affiliation{Texas Center for Superconductivity, University of Houston, Houston,
Texas 77204}
\date{\today}
\begin{abstract}

The Fulde-Ferrell-Larkin-Ovchinnikov (FFLO) states for two-dimensional $s$- and
$d$-wave superconductors (s- and d-SCs) are self-consistently studied under an
in-plane magnetic field. While the stripe solution of the order parameter (OP)
is found to have lower free energy in s-SCs, a square lattice solution appears
to be energetically more favorable in the case of d-SCs. At certain symmetric
sites, we find that the features in the local density of states (LDOS) can be
ascribed to two types of bound states. We also show that the LDOS maps for d-SCs
exhibit bias-energy-dependent checkerboard patterns. These characteristics can
serve as signatures of the FFLO states.
\end{abstract}
\maketitle

The inhomogeneous superconducting state known as the Fulde-Ferrell-Larkin-Ovchinnikov
(FFLO) state was predicted in the mid-1960's for a superconductor (SC) in a strong
exchange field, due to ferromagnetically aligned impurities~\cite{FFLO}. When
a high magnetic field ($H$) is applied parallel to the layers of a quasi-two-dimensional
(q-2D) SC, the Zeeman effect dominates over the orbital effect. Then the FFLO
state can also have lower free energy than a homogeneous superconducting
state~\cite{maki66, gru66}.

A number of q-2D High-$T_c$ SCs and q-1D and -2D organic SCs have been suggested
to have the FFLO phase in the low-temperature ($T$) and high ($H$)
limit~\cite{Buzdin,Tanatar,Balicas,Shimahara02,Houzet, Singleton,Burkhardt}.
Recent interest is on the q-2D heavy fermion compound CeCoIn$_5$, which
becomes a $d$-wave SC (d-SC) below a $T_c\sim$
2.3K~\cite{Petrovic,Movshovich,Izawa}. With a
strong $H$ applied parallel to its conducting planes, heat capacity
measurements reveal a second order phase transition into a new superconducting
phase, which strongly suggests that the system may have realized the FFLO
phase~\cite{Bianchi03, Radovan}. Various other measurements
~\cite{Capan,Watanabe, martin,kakuyanagi} also
support the existence of the FFLO state.

The direct observation of superconducting order parameter (OP) modulation is a
challenging task. The quasiparticle local density of states (LDOS) is a useful
quantity to measure for this purpose. It could be directly detected in scanning
tunneling microscopy (STM), which measures the local tunneling conductance between
a normal-metal tip and a SC. Vorontsov {\it et al.}~\cite{Vorontsov} have calculated the LDOS of 2D d-SC by solving quasi-classical Eilenberger
equations, but only 1D OP modulations have been considered.
References [21] and [22] treated the OP as a small parameter near the phase boundary, and predicted for d-SC that the energetically favored 
state is a square lattice, but they did not investigate the LDOS. In this letter, using a tight-binding
model, we have self-consistently determined the spatial variation of the FFLO-state OP in the low $T$ limit, for both 2D s-SCs and d-SCs, with an $H$ applied parallel to the conducting plane. We have then calculated the magnetization density and the LDOS. Our results show that for 2D s-SCs (d-SCs), 1D-stripes (2D-lattice) solutions are more energetically favorable. As a signature of 1D-stripes solutions for s-SCs, the LDOS spectrum shows two low-energy peaks on top of a low-energy bump, which indicate a midgap-states~\cite{hu94}-formed
mini-band with higher LDOS near the edges. In addition to two low-energy peaks arising from similar midgap-states, LDOS spectrum for d-SCs shows two additional low-energy peaks, which are outside the other two peaks in energy, and originate from finite-energy Andreev bound states (ABS) localized around saddle points of the OP, providing definitive signatures of 2D modulations in the OP.
Measuring fixed-bias spatial maps at these peak energies can confirm this interpretation.

We begin with a model system on a 2D square lattice with a pairing interaction
$V$ between two electrons on the same site for s-SCs, or on the nearest-neighbor
sites for d-SCs, which in the mean-field approximation leads to the discrete Bogoliubov-de-Gennes (BdG) equations:
\begin{equation}
\sum_j\left({H_{ij,\sigma}\atop \Delta_{ij}^*}{\Delta_{ij}\atop{-H_{ij,\bar\sigma}^*}}
\right)
\left(u_{j\sigma}^n\atop v_{j\bar\sigma}^n\right) =
E_n\left(u_{j\sigma}^n\atop v_{j\bar\sigma}^n \right),
\end{equation}
where the single particle Hamiltonian
$H_{ij,\sigma} = -t_{ij}-(\mu+\sigma h)\delta_{ij}$ with $\mu$ the chemical
potential, $h$ the Zeeman energy or the exchange interaction, and $t$ the effective hopping integral between two nearest-neighbor sites.
$(u_{j\sigma}^n,v_{j\bar\sigma}^{n})$ are the Bogoliubov quasiparticle amplitudes
on the $j$-th site. The superconducting OP satisfies the self-consistency condition
$\Delta_{ij} = \frac{V}{4}\sum_{n}(u_{i\uparrow}^nv_{j\downarrow}^{n*} +
u_{j\uparrow}^nv_{i\downarrow}^{n*})\tanh(E_n/2k_BT)$.
The s-SC OP at site $i$ is $\Delta_{ii}$ and the d-SC OP
at site $i$ is defined as $\Delta_i = (\Delta_{i,i+e_x} +
\Delta_{i,i-e_x}-\Delta_{i,i+e_y}-\Delta_{i,i-e_y})/4$, where $e_{x,y}$ denotes
the unit vector along $x,y$ direction.  For a 2D system with the magnetic field parallel to the superconducting plane, there is no orbital effect. (That is, the Doniach phase coupling~\cite{Doniach} between neighboring layers of q-2D layered superconductors is neglected.)
In case there are more than one solutions for a set of parameters, we compare
their free energies~\cite{Song} in order to find the energetically favored state.
We apply this method in real
space to study the FFLO state with a real OP.
A weak point of our approach is that we are unable to obtain solutions with
periods not commensurate with the lattice size. That the lattice size used here
can not be too large prevents us from obtaining solutions for small 
$h$ and 
the complete phase diagram. The advantage of our approach is
that it allows us to obtain the LDOS accurately and the physics associated with the spatial structure of the LDOS can be understood in terms of quasiparticle ABSs.

\begin{figure}
    \centering
    \includegraphics[width=2.5in,height=3in]{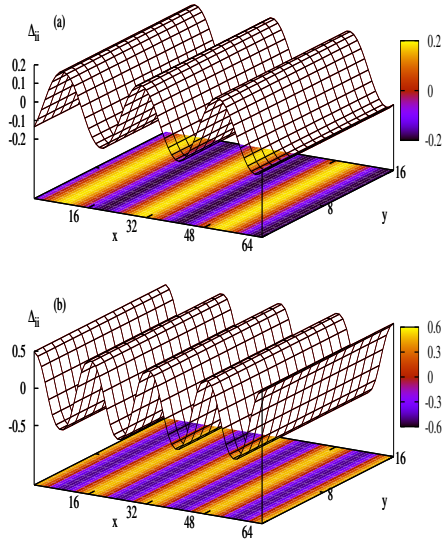}
    \caption{ Plots of the OP $\Delta_{ii}$ as a function of
    position for an s-SC. For (a), $V=2.0$, $h=0.25$;
    for (b), $V=2.5$, $h=0.4$. Here the chemical potential $\mu=-0.4$.}
    \label{fig:orders}
\end{figure}

The OP is calculated self-consistently by solving the BdG equations iteratively,
with the OP initially uniform on all sites. The calculation is made on
$16\times 16$ supercells each of which is a $32a\times 32a$ lattice and
$8\times32$ supercells each of which is a $64a\times 16a$ lattice, with
$a$ the lattice constant. The value of the OP at $H = 0$ is
determined by the chemical potential $\mu$ and the pairing
interaction $V$. The modulation period of the OP is close to
$\frac{\pi\hbar v_f}{h}$~\cite{kakuyanagi}, where $v_f$ is the Fermi velocity,
which is determined by $\mu$. 
Thus one can adjust the period by changing either $\mu$ or $h$. To change $h$,
we have to change $V$ as well, because an FFLO state exists only for $h$ in
an interval $[\mu_0 H_{c1},\mu_0 H_{c2}]$~\cite{FFLO}, where $\mu_0$ is the
magnetic moment of an electron, and the lower and upper critical fields $H_{c1}$
and $H_{c2}$ are both proportional to the superconducting gap at zero $H$. In
order to obtain solutions that are commensurate with the lattice, we relax the
restriction on $\mu$ and $V$ from the values appropriate to any specific material.
Throughout our calculation, we let $t = 1$ and $T = 0.001$.
\begin{figure}
    \centering
    \includegraphics[width=2.5in,height=3.0in]{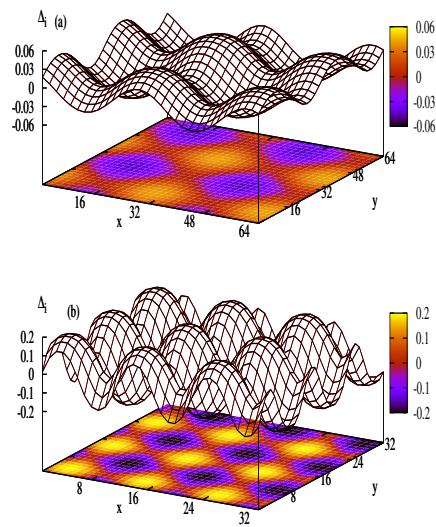}
    \caption{Plots of the OP $\Delta_i$ as a functions of position
    for a d-SC. For (a), $V=1.0$, $h=0.15$ and $\mu=-0.4$;
    for (b), $V=2.0$, $h=0.4$ and $\mu=-0.4$.}
    \label{fig:orderd}
\end{figure}

The energetically favored solution in s-SCs always forms 1D stripes, which fits closely to the form $\Delta_0\cos(qx)$, with 
$\Delta_0$ the maximum value of the OP which is not necessarily
the value of the OP at zero field.  We use a $64a\times16a$ lattice 
to obtain longer period solutions. Figure \ref{fig:orders}
shows the OP $\Delta_{ii}$ as a function of position for two
choices of $V$ and $h$ at $\mu=-0.4$ for s-SCs. The 1D modulation 
of the OP with a periodicity $64a/3$ is plotted in Fig.~\ref{fig:orders}(a) with $V=2.0$ and $h=0.25$. Using $V=2.5$, we see in Fig.~\ref{fig:orders}(b) that
the periodicity becomes $16a$ at $h=0.4$. We find that the periodicity decreases
from $16a$ to $64a/5$ as the Zeeman energy increases from $h=0.4$ to 0.45. For
$V=2.0$ and $V=2.5$, the OPs at zero field are $0.34t$ and $0.57t$ respectively.
The nodal lines where the OP changes sign are along (010) or the y
direction.

The OP of the energetically favored states in d-SCs always forms a square lattice,
which fits very closely to the form of $\Delta_0[\cos(qx)+\cos(qy)]$, as shown in
Fig.~\ref{fig:orderd}. As a consequence, the nodal lines
of the OP are in (110) and (1$\bar 1$0) directions. The lattice used for our
numerical calculations is $32a\times 32a$. 1D modulation solutions for d-SCs
by using a $64a\times16a$ lattice are also obtained, but they have higher
free energy per site than 2D lattice solutions on a $32a\times32a$ lattice.
Fig.~\ref{fig:orderd}(a) shows the profile of the OP at $V=1.0$.
Here, the OP at zero field is $0.057 t$. In order to show the periodic
structure of spatial profiles more clearly, for this set of
parameters, we have plotted our results on a $64a\times 64a$ lattice by spatial
repetition. Fig.~\ref{fig:orderd}(b) presents the profile of the OP at $V=2.0$.
For this $V$, the OP at zero field is $0.23 t$.
\begin{figure}
\centering
\includegraphics[width=2.5in,height=2.66in]{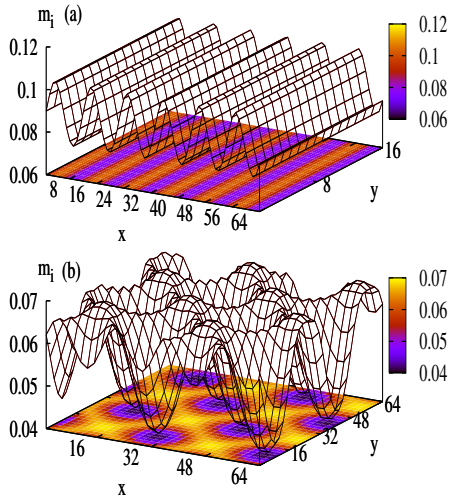}
\caption{The spatial profile of magnetization $m_i$ of the FFLO state. (a) is for an
s-SC with $V=2.0$, $h=0.25$ and $\mu$=-0.4. (b) is for a d-SC with $V=1.0$, $h=0.15$
and $\mu=-0.4$.}
\label{fig:mag}
\end{figure}

The magnetization $m_i$=$n_{i\uparrow}$-$n_{i\downarrow}$, with $n_{i\uparrow}$
($n_{i\downarrow}$) as the number of spin-up (-down) electrons at site $i$,
has modulations shown in Fig.~\ref{fig:mag}. The parameters of Fig.~\ref{fig:mag}
(a) are the same as those in Fig.~\ref{fig:orders} (a).
The parameters of Fig.~\ref{fig:mag} (b) are the same as those in
Fig.~\ref{fig:orderd} (a). The period of modulation here is one half of
that of the OP. The magnetization in an s-SC (a d-SC) forms 1D stripes (2D lattice).
The magnitude of the magnetization is highest at the nodal lines of the OP.

The LDOS of spin-up and -down quasi-particles can be calculated from
\begin{equation}
\rho_{i\sigma}(E)=\sum_n [|u^n_{i\sigma}|^2\delta(E_n-E) +
|v^n_{i\bar\sigma}|^2\delta(E_n+E)]
\end{equation}
where the delta function $\delta(x)$ has been approximated by
$\epsilon/\pi(x^2+\epsilon^2)$. $\epsilon$ is set to 0.01 in our
calculation.

\begin{figure}
\centering
\includegraphics[width=3.0in,height=2.4in]{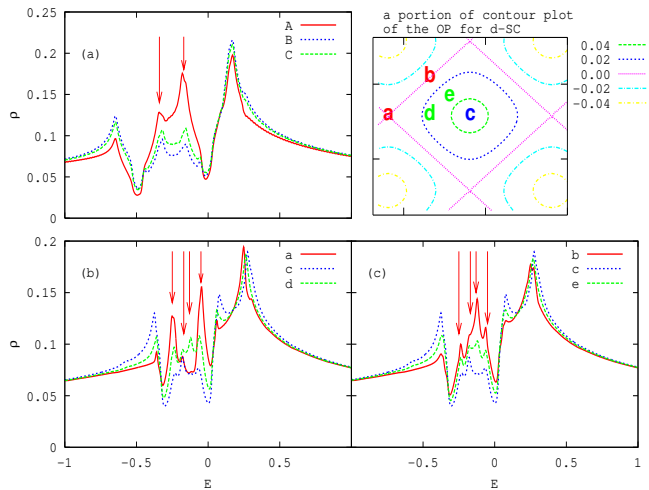}
\vskip 0.2in
\caption{The spin-up LDOS spectrum in the FFLO state. (a): the LDOS at three
different locations (A, B, and C, see text,) for an s-SC with $V=2.0$,
$h=0.25$ and $\mu=-0.4$. In the upper-right corner, the contour plot of the
OP for a d-SC is displayed with five symmetric sites. (b) and (c): Plots
similar to (a) for a d-SC with $V=1.0$, $h=0.15$ and $\mu=-0.4$ at sites a,
b, c, d and e (see text). The locations of the LDOS peaks are marked by arrows.}
\label{fig:ldos}
\end{figure}

In Fig.~\ref{fig:ldos}(a), we plot the spin-up LDOS for an s-SC at three
sites with reference to Fig.~\ref{fig:orders}(a). Curve A is for a site on
the nodal line, curve B is for a site where $|\Delta_{ii}|$ has the maximum
absolute value, and curve C is for a midpoint between sites in A and B. In
the upper-right corner of  Fig.~\ref{fig:ldos}, we have the contour plot of
the OP for a d-SC with reference to Fig.~\ref{fig:orderd}(a). Here a is the
saddle point where two nodal lines intersect, b is the mid-site between
two neighboring saddle points, c is the site where the OP has the maximum
magnitude, and d and e are respectively the mid-points between a-c and b-c.
The spin-up LDOS at these symmetric sites are shown in Figs.~\ref{fig:ldos}(b)
and (c). The spin-down LDOS spectra (not shown) are simply
the curves in Fig.~\ref{fig:ldos} shifted to the right by $2h$.
Notice that there is a van Hove peak located at $E=0.17$ ($E=0.27$) in
Fig.~\ref{fig:ldos}(a) [Figs.~\ref{fig:ldos}(b) and \ref{fig:ldos}(c)].
The LDOS spectrum is skewed by this van Hove peak. These plots reveal that
there are two kinds of ABSs in the d-wave case: One due to the sign
change of the OP across the nodal lines, and is similar to the midgap
states~\cite{hu94}. These states would have essentially zero energy if not
for the Zeeman shifts to $\mp h$ for spin-up and -down quasiparticles,
and broadening into a mini-band due to
tunneling between the different nodal lines in a periodic FFLO state. This
feature already exist in an s-SC, as shown in Fig.~\ref{fig:ldos}(a), and is
the signature of a 1D FFLO state. The semiclassical orbit of such a state
is a round trip along a straight-line segment across a nodal line, ending
with two Andreev reflections involving opposite signs of the OP
on the two sides of the nodal line. The same mechanism also gives rise to the
inner two peaks in Figs.~\ref{fig:ldos}(b) and \ref{fig:ldos}(c) which are for
a d-SC. The second kind of ABSs is essentially localized at the saddle
points of the OP. Because the FFLO state in d-SC is a 2D lattice, the
absolute-OP is very much suppressed in a cross-shaped region at every saddle
point, which yields a potential well for a quasiparticle and thus generates
two finite-energy ABSs (per spin) at every saddle point. Zeeman shifts
also apply to these states, and communication between the saddle points turns
these states into two narrow mini-bands. The semiclassical orbits of these
states are round trips along straight line segments across the saddle points,
ending with two Andreev reflections involving only one sign of the OP.
Thus these are not related to the midgap states, and should not have zero
energy before Zeeman shift and broadening. The LDOS plot at a saddle point
[the curve a in Fig. \ref{fig:ldos}(b)]
exhibits two strong peaks corresponding to two ABSs located at
$-h\pm\epsilon_0$ with $\epsilon_0 = 0.10$, where $-h$ is the Zeeman shift.
One can see that curve a is modified by two weak-inner peaks originated
in the midgap states. Because of non-locality of these states, 4 low-energy
peaks show up at various symmetric points in Figs.4(b) and 4(c).  They
are located at $E = -0.25$, $-0.17$, $-0.13$ and $-0.05$. (Removing the Zeeman
shift there would be at $E = \pm 0.10$ and $\pm 0.02$, representing two $\pm E$
pairs of ABSs of different nature.)
\begin{figure}
\centering
\includegraphics[width=3.0in,height=1.8in]{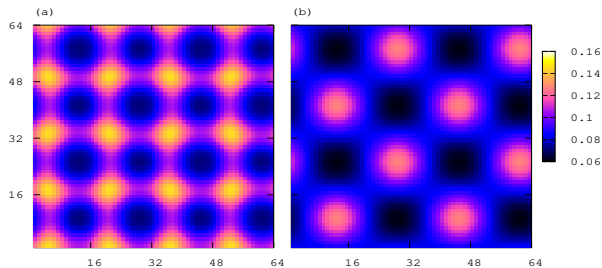}
\caption{The LDOS maps for spin-up quasiparticles at certain
low-energy-peak biases $E$ for d-SC with $V = 1.0$, $h=0.15$
and $\mu=-0.4$. For (a), $E=-0.13$. For (b), $E=-0.25$.}
\label{fig:ldos0}
\end{figure}

For an s-SC, the LDOS map for spin-up quasiparticles at either of the two
low-energy peaks, not presented here, forms simple 1D stripes and its
intensity is highest at the nodal lines of the OP, as is expected for
zero-energy ABS (or midgap states) formed near the nodal lines.

For a d-SC, the structure of the spin-up LDOS {\it maps} at the four
low-energy-peak biases are richer in physics, and are presented in
Figs.~\ref{fig:ldos0}(a) and \ref{fig:ldos0}(b). Fig.~\ref{fig:ldos0}(a)
shows the LDOS map at $E=-0.13$ (with similar result at $E=-0.17$.)
The intensity is highest at points on nodal lines halfway between two
neighboring saddle points, and is lowest at both the saddle points and
the absolute-OP maxima. The high-intensity spots form a
$16a\times16a$ checkerboard pattern with unit vectors along the (100)
and (010) directions. Figure~\ref{fig:ldos0}(b) is at $E=-0.25$
(with similar result if $E=-0.05$).  The highest intensity is at saddle
points of the OP, and they form a $16\sqrt{2}a\times 16\sqrt{2}a$
checkerboard pattern but the unit vectors are along (110) and
$(1{\bar 1}0)$.  Thus the two peaks at $E=-0.17$ and $-0.13$ correspond
to zero-energy ABS formed near the nodal lines, similar to the case in the s-SC.
On the other hand, the two peaks at $E=-0.25$ and $-0.05$ correspond to
finite-energy ABSs formed near the saddle points.
Note that if the STM tip is unpolarized, then the measured LDOS
should come equally from the spin-up and spin-down contributions.
Even under this situation our calculation indicates that the LDOS
{\it maps} remain qualitatively unchanged as compared to Fig.~\ref{fig:ldos0}.
 
In Summary, using a tight-binding model, we have studied, in the low $T$
limit, the FFLO state in a 2D SC with a magnetic field applied in plane.
The superconducting OP is self-consistently determined. The spatial profile
of the energetically-favored solution is found to form 1D stripes for an s-SC,
and a 2D lattice for a d-SC. The spatially-varying magnetization and the
quasiparticle LDOS spectrum have been calculated. At all symmetric sites
of the FFLO lattice, we find that the low-energy features in the LDOS can
be ascribed to two types of ABSs, one is related to zero-energy ABSs
(or midgap states) and the other is not. LDOS maps are presented for a d-SC
at certain low-energy-peak biases. They are shown to have bias-dependent
checkerboard patterns which are different from the spatial profile of the OP.
Measuring such maps can clearly reveal the nature of the two different types
of ABSs. These characteristics provide clear signatures of the 2D FFLO state
in d-SCs. In case the phase of CeCoIn$_5$ has a competing antiferromagnetic order, we need to use a model suitable for the heavy fermion system to reexamine this problem.

This work is supported by a grant from the Robert A. Welch Foundation under
NO. E-1146.

\end{document}